\documentclass[final]{svjour3}
\usepackage{hyperref}
\usepackage{rotating}
\usepackage{amssymb}
\usepackage{mathptmx}
\usepackage[numbers]{natbib}
\makeatletter
\journalname{Journal of Low Temperature Physics}
\usepackage{color}

\begin{document}

\newcommand{\hdblarrow}{H\makebox[0.9ex][l]{$\downdownarrows$}-}
\title{Measurement of optical constants of TiN and TiN/Ti/TiN multilayer films for microwave kinetic inductance photon-number-resolving detectors}

\author{M. Dai$^1$, W. Guo$^2$, X. Liu$^3$, M. Zhang$^3$, Y. Wang$^3$, L. F. Wei$^1$, G. C. Hilton$^2$, J. Hubmayr$^2$, J. Ullom$^2$, J. Gao$^{2}$ \and M. R. Vissers$^2$}

\institute{1. Information Quantum Technology Laboratory, School of Information Science and Technology, Southwest Jiaotong University, Chengdu, 610031, China\\
2. National Institute of Standards and Technology, Boulder, CO 80305, USA\\
3. Quantum Optoelectronics Laboratory, School of Physical Science and Technology, Southwest Jiaotong University, Chengdu, 610031, China\\
\email: yiwenwang.nju@gmail.com, lfwei@swjtu.edu.cn}
\maketitle

\begin{abstract}
We deposit thin titanium-nitride (TiN) and TiN/Ti/TiN multilayer films on sapphire substrates and measure the reflectance and transmittance in the wavelength range from $400$~nm to $2000$~nm using a spectrophotometer. The optical constants (complex refractive indices), including the refractive index $n$ and the extinction coefficient $k$, have been derived. With the extracted refractive indices, we propose an optical stack structure using low-loss amorphous Si (a-Si) anti-reflective coating and a backside aluminum (Al) reflecting mirror, which can in theory achieve 100\% photon absorption at $1550$~nm. The proposed optical design shows great promise in enhancing the optical efficiency of TiN-based microwave kinetic inductance photon-number-resolving detectors.


\end{abstract}

\section{Introduction}
Photon-number-resolving (PNR) detectors are able to directly measure the photon number and energy in a pulse of incident light. In particular, the PNR detectors at visible and near-infrared wavelengths have important applications in many fields such as quantum secure communications~\cite{hiskett2006long}, linear optics quantum computing~\cite{knill2001scheme}, quantum optics experiments~\cite{Bell2013} and optical quantum metrology~\cite{zwinkels2010photometry}. To meet the requirements of these applications, an ideal PNR detector should have both high energy resolution and high system detection efficiency. By minimizing the fiber-to-detector coupling losses and using optical stack structures that enhance the photon absorption by the absorber material, transition edge sensors (TESs) have demonstrated high energy resolution and near unity system detection efficiency at near-infrared wavelengths~\cite{miller2003demonstration,lita2008counting,Lita10,calkins2013high,lolli2012ti,brida2012quantum,lolli2013high}.

Another type of superconducting detector with intrinsic photon-number-resolving and energy-resolving capability is the microwave kinetic inductance detector (MKID)~\cite{Day2003}. As compared to TESs, MKIDs are easy to fabricate and multiplex into large arrays. Single-photon counting at telecommunication wavelengths (near-infrared) with titanium-nitride (TiN) MKIDs was first demonstrated in reference~\cite{JG12}. Recently, by optimizing the MKID design, we have achieved an energy resolution of $0.22$~eV and resolved up to $7$ photons per optical pulse at $1550$~nm using MKIDs made from TiN/Ti/TiN trilayer films~\cite{WG17}. Although the energy resolution of these detectors is already impressive, little effort has been made to improve their optical efficiency.


The major sources that limit the optical efficiency of MKIDs include fiber-to-detector coupling efficiency, photon loss due to reflection from and transmission through the thin TiN layer. As the first step to improve the optical efficiency of photon-number-resolving TiN-MKIDs, in this paper, we show measurements of the optical constants (complex refractive indices) for thin TiN and TiN/Ti/TiN multilayer films. We then propose an optical stack structure that can in theory achieve 100\% photon absorption at $1550$~nm for TiN-based MKIDs.

\section{Principle}
In this section, we briefly introduce the optical transfer matrix theory which is commonly used to obtain the reflectance $R$ and transmittance $T$~\cite{Nestell72} for a multilayer optical structure.

Consider a beam of light from the air is normally incident onto an optical stack consisting of $K$ layers and exit into the air. Assume the refractive index of air is $1$ and each layer has thickness of $d_j$ and complex refractive index of $\tilde{n}_j= n_j - i k_j$ (here $n_j$ and $k_j$ are wavelength $\lambda$ dependent refractive index and extinction coefficient for the $j$-th layer, $j$ = 1, ..., K), the magnitude of the incident electromagnetic wave $E_0$ and the transmitted wave $E_{K+1}$ are related by the products of the transfer matrices as


\begin{eqnarray}
  \vec{E}_{0}\left[
         \begin{array}{c}
           1 \\
           Y \\
         \end{array}
       \right]=\prod_{j=1}^{K}\left[
                  \begin{array}{cc}
                    \cos(k\tilde{n}_{j}d_{j}) & i\cdot\sin(k\tilde{n}_{j}d_{j})/\tilde{n}_{j} \\
                    i\cdot \tilde{n}_{j}\sin(k\tilde{n}_{j}d_{j}) & \cos(k\tilde{n}_{j}d_{j}) \\
                  \end{array}
                \right]\left[
                         \begin{array}{c}
                           1 \\
                           1 \\
                         \end{array}
                       \right]\vec{E}_{K+1}=\left[
                                                         \begin{array}{c}
                                                           B \\
                                                           C \\
                                                         \end{array}
                                                       \right]\vec{E}_{K+1},\label{eqn:tmatrix}
\end{eqnarray}
where $[B~~C]^{T}$ is called the characteristic matrix of dielectric films and $Y=C/B$ is defined as the optical admittance. It follows from Eq. (1) that the reflectance and transmittance of the whole optical stack are given by
\begin{eqnarray}
  R(\lambda, n_j, k_j, d_j) &=& \left|\frac{B-C}{B+C}\right|^{2},\\
  T(\lambda, n_j, k_j, d_j) &=& =\left|\frac{2}{B+C}\right|^{2},
\end{eqnarray}

Eq. (2-3) allow us to solve for the $n$ and $k$ values of a layer in the optical stack from the combined measurements of the reflectance $R$ and transmittance $T$, if we know the complex refractive indices for all the other layers and the thickness of all the layers. In Section $3$, we use this method to extract the complex refractive indices for thin TiN and TiN/Ti/TiN multilayer films deposited on thick sapphire substrates. On the other hand, when the complex refractive indices $n_j-ik_j$ and thickness $d_j$ of each layer are given, Eq. (2-3) also allow us to directly calculate the wavelength dependent $R(\lambda)$ and $T(\lambda)$ of the multilayer stack, which is used in Section $4$ to design the optimized optical structure that can maximize the photon absorption at $1550$~nm.


\section{Measurements of optical constants}

We deposited thin stoichiometric TiN and TiN/Ti/TiN multilayer films on $500$~$\mu$m thick double-side polished sapphire substrates. The multilayer film comprises a stack of 7 layers with TiN at the top and bottom interface and alternating layers of Ti and TiN in the middle. The detailed deposition conditions and processes can be found in reference~\cite{Vissers10, Vissers2013}. To precisely determine the optical constants of these thin films, one should properly choose the film thickness. On one hand, if the film is too thin, the uncertainty in the film thickness will result in large errors in the derived $n$ and $k$ values. On the other hand, if the film is too thick, the transmitted light will be too weak to be measured accurately due to instrument noise and background stray light. We have chosen a target thickness of 60~nm for both films. The actual thicknesses were carefully measured by a profilometer which reports $63$~nm for the stoichiometric TiN film and $43$ nm for the TiN/Ti/TiN multilayer.

The reflectance $R$ and transmittance $T$ in the wavelength range from $400$~nm to $2000$~nm were measured at room temperature using a commercial spectrophotometer (PerkinElmer's LAMBDA 1050), and the results are shown in Fig.~1(a) and (b), respectively.  From the $R$ and $T$ data, it is straightforward to obtain the absorption as $A = 1-R-T$, which is of practical interest since it directly affects the optical efficiency.

\begin{figure}[ht]
	\centering
	\includegraphics[width=11.2cm,height=4.23cm]{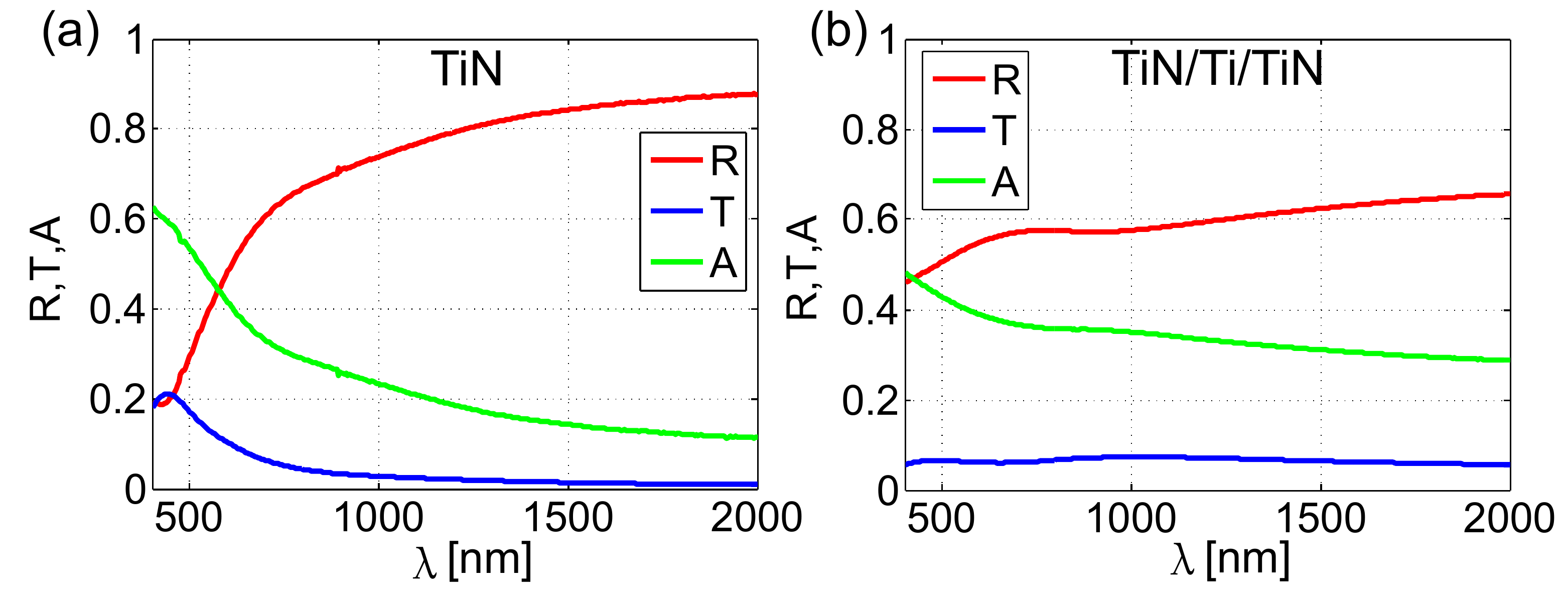}
		\caption{Measurements of the reflectance $R$ and transmittance $T$ for: (a) $63$~nm stoichiometric TiN film on a $500$~$\mu$m-thick sapphire substrate. (b) $43$~nm TiN/Ti/TiN multilayer film on a $500$~$\mu$m-thick sapphire substrate.}
	\label{Fig.1}
\end{figure}

We also measured the $R$ and $T$ data for a $500$~$\mu$m thick double-side polished bare sapphire wafer, which can be easily analyzed with the transfer matrix theory introduced in Section 2 to give the complex refractive index of sapphire ($n_2 = 1.75$ and $k_2=0$), which is approximately wavelength independent. We find the measured optical constants of sapphire match well with the tabulated values~\cite{sapphire}. For the two-layer stack of TiN film on sapphire substrate, we have the measured thickness of the TiN layer ($d_1 = 63$~nm for stoichiometric TiN and $43$~nm for TiN/Ti/TiN multilayer), the vendor specified wafer thickness ($d_2 = 500~\mu$m) and the derived complex refractive index ($n_2 = 1.75$ and $k_2=0$) of the sapphire substrate. Then Eq. (2-3) reduce to two nonlinear equations with two unknown variables ($n_1$ and $k_1$ of the TiN layer),
\begin{eqnarray}
  \left\{
        \begin{array}{lll}
        R(n_1, k_1; \lambda, n_2, k_2, d_2) &=& R_\mathrm{mea}\\
        T(n_1, k_1; \lambda, n_2, k_2, d_2) &=& T_\mathrm{mea}\\
        \end{array}
        \right.,
\end{eqnarray}
where $R_\mathrm{mea}$ and $T_\mathrm{mea}$ are the measured reflectance and transmittance data. The above equations can be solved numerically~\cite{Nestell72} at a specific wavelength $\lambda$. Note that we take the thick-substrate approximation~\cite{rosenberg2004near} that assumes reflections off the back surface add incoherently, so that we can get smooth $n$ and $k$ vs. wavelength curve.


Fig.~2(a) and (b) plot the derived refractive index $n$ and extinction coefficient $k$ as a function of wavelength. For stoichiometric TiN film, one can see a low refractive index in the visible wavelength and even $n < 1$ at the wavelength range from $570$~nm to $712$~nm, which is similar to previous optical measurements on TiN films~\cite{Valkonen86,karlsson1982optical}. The $n$ for TiN/Ti/TiN multilayer is higher and increases monotonically with wavelength, which resembles the properties of pure Ti film. $k$ ranges from $2$ to $6$ is also a signature of pure Ti film~\cite{sapphire}.


\begin{figure}[ht]
        \centering
		\includegraphics{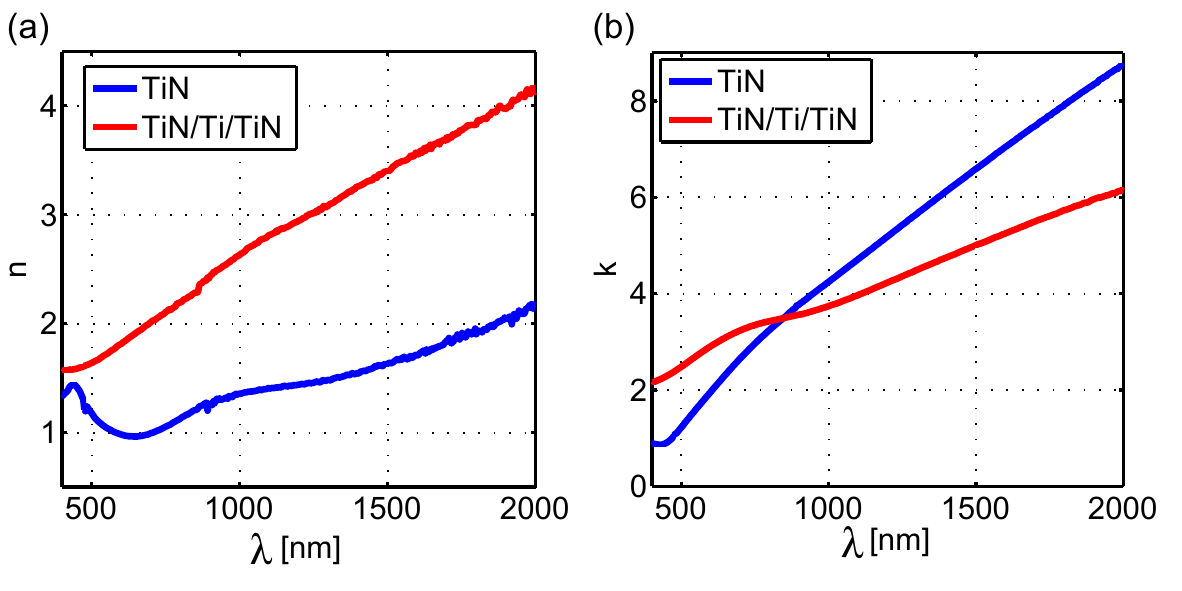}
		\caption{The derived optical constants ($n$ and $k$) vs. wavelength. The blue and red curves represent stoichiometric TiN and TiN/Ti/TiN multilayer films, respectively.}
\end{figure}

\section{Optical designs for enhancement of photon absorption}

In this Section, we use the $n$ and $k$ values derived in the previous section to design optimal stack structures that can enhance the photon absorption efficiency at a particular wavelength. In the following we mainly discuss the designs for TiN/Ti/TiN multilayers at $1550$~nm, because photon-number-resolving MKIDs based on TiN/Ti/TiN multilayer films have already demonstrated high energy resolution and good multiphoton discrimination capability at $1550$~nm~\cite{WG17}, and also because the TiN/Ti/TiN multilayer has superior uniformity (as compared to substoichiometric TiN films) which is critical for scaling into large MKID arrays.

\begin{figure}[ht]
	\centering
	\includegraphics{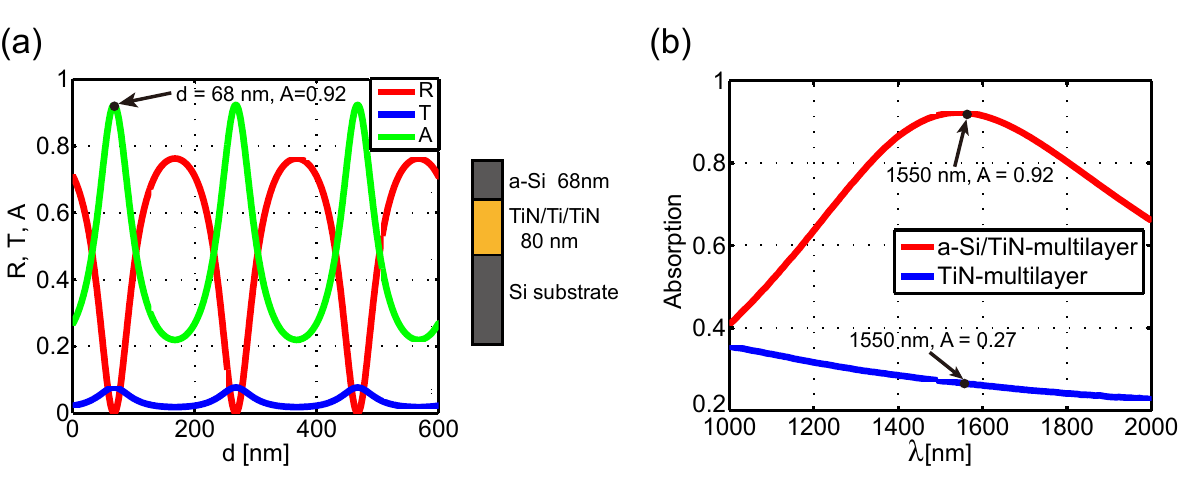}
		\caption{(a) The simulated reflectance $R$, transmittance $T$ and absorption $A$ at $1550$~nm of AR coated TiN/Ti/TiN multilayer, showing a periodical change as a function of the thickness of a-Si AR layer. With a $68$~nm thick a-Si AR layer, the absorption can reach a maximum of 92~\%. The stack structure as well as the thicknesses of each layer are illustrated by the small cartoon to the right. (b) Comparison of the absorption $A$ by $80$~nm TiN/Ti/TiN multilayer with (red curve) and without  anti-reflective coating (blue curve) in the near-infrared wavelength, which clearly show the AR coating can significantly improve the absorption.}
\end{figure}

We present two optical designs for enhancing photon absorption in TiN/Ti/TiN multilayer films. The first relative simple structure includes a single amorphous Si (a-Si) anti-reflective (AR) coating layer deposited on top of a $80$~nm thick TiN/Ti/TiN multilayer film. We choose a-Si because it has shown reduced two-level systems (TLSs) and low microwave loss comparing to other dielectric materials, such as SiO$_2$ and Si$_3$N$_4$~\cite{martinis2005decoherence}. We choose a relative thick ($80$~nm) TiN/Ti/TiN multilayer film to enhance the absorption in the film and prevent light from penetrating the film. Using the extracted values of $n$ and $k$, we calculate expected performance of this structure in terms of reflectance, transmittance and absorption at $1550$~nm wavelength as a function of the thickness of a-Si AR coating. As shown in Fig.~3(a), the absorption varies periodically with the thickness of Si layer and a maximum absorption of $92~\%$ is first reached at a coating thickness of $68$~nm, which suggests an optimal design illustrated by the cartoon in Fig.~3(a). Fig.~3(b) shows the simulated absorption of this optimal stack (red curve) between $1000$~nm to $2000$~nm, which is compared to the TiN/Ti/TiN multilayer film alone without AR coating (blue curve).  It is clear that the AR coating design has significantly enhanced the absorption, from less than 30\% to over 90\% around $1550$~nm.

While the single-layer a-Si AR coating solution is attractive and relative easy to implement (and we plan to test it in the future), the theoretical maximum absorption is still less than unity, so we further propose another optical stack design that can achieve 100\% absorption at 1550nm. In this design, a thin TiN/Ti/TiN trilayer film is sandwiched between two a-Si layers of different thicknesses and a thick aluminum (Al) layer is buried under the sandwich. Different from the first design, the second design accommodates a thin $20$~nm TiN/Ti/TiN trilayer film, which is the film used in our previous photon-counting experiment \cite{WG17}. In that experiment, we found that absorbers with smaller volume showed larger responsivity and better energy resolution. We use an Al layer as a bottom mirror to create an optical cavity and embed the thin TiN/Ti/TiN trilayer film in the cavity to enhance the absorption. An important reason for choosing Al (instead of Au as used in optical TES) as the mirror material is that Al is superconducting which is expected not to introduce metal losses and degrade the quality factor of the MKIDs.

Fig.~4(a) shows the simulated absorption at $1550$~nm in the two-dimensional parameter space of the two a-Si layer thicknesses ($d_1$ and $d_2$). The absorption varies periodically with $d_1$ and $d_2$ and a maximum absorption of unity is first reached at $d_1 = 56$~nm and $d_2 = 35$~nm. The stack structure with optimal layer thicknesses is illustrated by the cartoon on the right of Fig.~4(a). In Fig.~4(b), we plot the wavelength-dependent absorption of this optimal stack structure, which shows that 100\% absorption and 0\% reflectance have been achieved at the wavelength of $1550$~nm.

\begin{figure}[ht]
	\centering
	\includegraphics{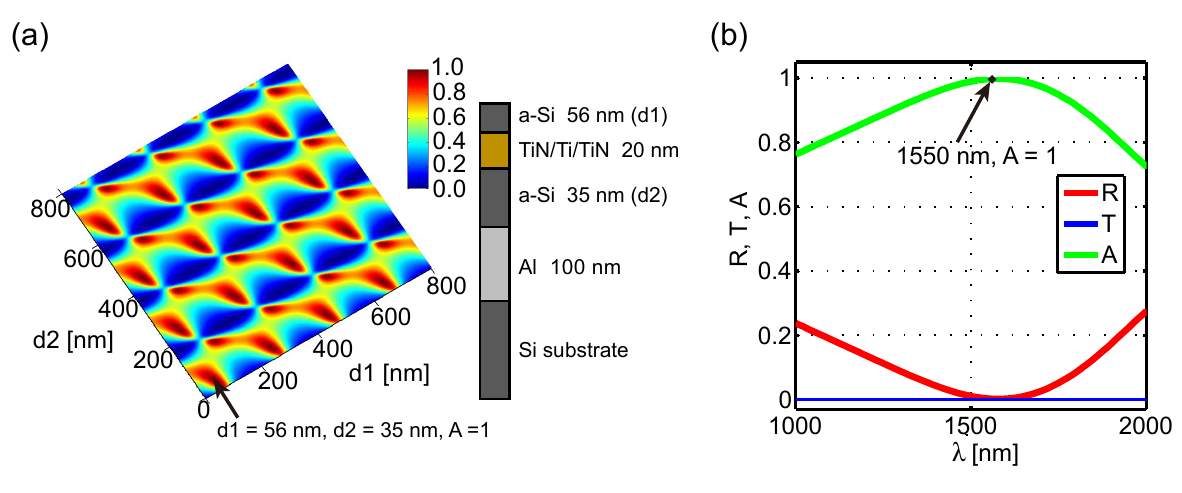}
		\caption{(a). The simulated absorption of a $20$~nm TiN/Ti/TiN trilayer vs. thicknesses of the top ($d_1$) and bottom ($d_2$) a-Si layers in the proposed optical cavity structure. The stack structure as well as the optimal thickness of each layer are illustrated by the small cartoon to the right. (b). The simulated wavelength-dependent reflectance $R$, transmittance $T$ and absorption $A$ of TiN/Ti/TiN multilayer film embedded in the optimized optical cavity structure.}
\end{figure}




\section{Conclusion}
We have successfully determined the optical constants (refractive index $n$ and extinction coefficient $k$) of superconducting stoichiometric TiN and TiN/Ti/TiN multilayer thin films, from combined measurements of reflectance and transmittance at the wavelength ranging from $400$~nm to $2000$~nm. By utilizing the extracted $n$ and $k$ values, two optical stack designs have been presented and discussed, which may significantly improve photon absorption into TiN/Ti/TiN multilayer films. In particular, by embedding the TiN/Ti/TiN film in an optical cavity structure one may in theory achieve unity absorption at $1550$~nm. The proposed designs show great promise in achieving high system detection efficiency in near-infrared photon-number-resolving TiN MKIDs.

\begin{acknowledgements}
The TiN films were deposited in the NIST-Boulder micro-fabrication facility. We thank Dr. Adriana. E. Lita for useful discussions. This work was supported in part by the National Natural Science Foundation of China (Grant Nos. 61301031, U1330201).
\end{acknowledgements}


\bibliography{references}
\bibliographystyle{JLTPv2}%

\end{document}